\documentclass{aa}
\usepackage{graphics}
\usepackage{psfig}

\begin{document}

\newcommand{\al}    {\rm et al.}
\newcommand{\eg}    {\em e.g.}
\newcommand{\ie}    {\em i.e.}
\newcommand{\met}   {metallicity }
\newcommand{\kms}   {km s$^{-1}$}
\newcommand\beq{\begin{equation}}
\newcommand\eeq{\end{equation}}
\newcommand\beqa{\begin{eqnarray}}
\newcommand\eeqa{\end{eqnarray}}
\newcommand{\aaa} [2]{A\&A {\bf #1}, #2}
\newcommand{\aas} [2]{A\&A Suppl. {\bf #1}, #2}
\newcommand{\aj}  [2]{AJ {\bf #1}, #2}
\newcommand{\apj} [2]{Ap. J. {\bf #1}, #2}
\newcommand{\apjl}[2]{Ap. J. Letter {\bf #1}, #2}
\newcommand{\apjs}[2]{Ap. J. Suppl. {\bf #1}, #2}
\newcommand{\araa}[2]{A\&AR {\bf #1}, #2}
\newcommand{\pasp}[2]{PASP {\bf #1}, #2}
\newcommand{\mnras}[2]{MNRAS {\bf #1}, #2}


\title{Exploration of the BaSeL stellar library for 9 F-type stars COROT potential targets$^{\star}$} 
\subtitle{Comparisons of fundamental stellar parameter determinations.} 

\author{E. Lastennet \inst{1,2}, F. Ligni\`eres \inst{1,3}, R. Buser
\inst{4}, Th. Lejeune \inst{5}, \\ 
Th. L\"uftinger \inst{6}, F. Cuisinier
\inst{2} and C. van 't Veer-Menneret \inst{7}
}

\institute{Astronomy Unit, Queen Mary and Westfield College,
           Mile End Road, London E1 4NS, UK
     \and  Observat\'orio do Valongo, Depto de Astronomia, UFRJ, 
           Ladeira Pedro Ant\^onio 43, 20080-090 Rio de Janeiro RJ, Brazil 
     \and  Laboratoire d'Astrophysique, CNRS UMR 5572, 
           Obs. Midi-Pyr\'en\'ees, 
           57 avenue d'Azereix, 65008 Tarbes Cedex, France              
     \and  Astronomisches Institut der Universit\"at Basel, Venusstr. 7,     
           CH-4102 Binningen, Switzerland              
     \and  Observat\'orio Astron\'omico de Coimbra, Santa Clara,              
           P-3040 Coimbra, Portugal       
     \and  Institut f\"ur Astronomie, Universit\"at Wien, 
           T\"urkenschanzstr. 17, A-1180 Wien, Austria  
     \and  Observatoire de Paris, 61 avenue de l'Observatoire, 
           F-75014 Paris, France
}

\offprints{E. Lastennet \\
$\star$ Based on observations made on the 193cm telescope at Observatoire de 
Haute-Provence, France and data from the ESA Hipparcos satellite}

\date{Received 13 April 2000 / Accepted 20 October 2000}

\authorrunning{E. Lastennet {\al}}
\titlerunning{Exploration of the BaSeL stellar library for 9 F-type 
stars COROT potential targets} 


\abstract{
The Basel Stellar Library (BaSeL models) is constituted of the merging of various 
synthetic stellar spectra libraries, with the purpose of giving the most comprehensive 
coverage of stellar parameters. It has been corrected for systematic deviations 
detected in respect to UBVRIJHKLM photometry at solar metallicity, and can then be 
considered as the state-of-the-art knowledge of the broad band content of stellar 
spectra. 
In this paper, we consider a sample of 9 F-type stars with detailed spectroscopic 
analysis to investigate the Basel Stellar Library in two photometric systems 
simultaneously, Johnson (B$-$V, U$-$B) and Str\"omgren (b$-$y, m$_1$, and c$_1$). 
The sample corresponds to potential targets of the central seismology programme
of the COROT space experiment, which have been recently observed at OHP.
The atmospheric parameters T$_{\rm eff}$, [Fe/H], and log g obtained from the BaSeL 
models are compared with spectroscopic determinations as well as with results of other 
photometric calibrations. 
For a careful interpretation of the BaSeL solutions, we computed confidence regions 
around the best $\chi$$^2$-estimates and projected them on T$_{\rm eff}$-[Fe/H], 
T$_{\rm eff}$-log g, and log g-[Fe/H] diagrams. We first derive the 3 atmospheric 
parameters from the full photometric information available (Johnson and Str\"omgren data).  
The BaSeL library has only been calibrated for broad band UBVRIJHKLM 
photometry, and it presents therefore intrinsic limitations in respect to other 
photometric systems, especially with different bandwidth.
Thus, using this combination (Johnson and Str\"omgren), BaSeL temperatures are 
systematically lower ($\sim$ 130K), and the discrepancy for the gravity and the 
metallicity can be quite large in comparison to the other methods. 
To disentangle these unexpected discrepancies, we fixed the temperatures at their 
spectroscopic values and studied the relative influence of each colour index in 
the log g-[Fe/H] diagrams. 
We find inconsistent results between Str\"omgren and Johnson synthetic colours. 
While the Johnson colours give impressively good log g-[Fe/H] solutions, the combination 
of the m$_1$ and c$_1$ Str\"omgren synthetic indices does not provide reliable results. 
This is not, of course, due to any intrinsic superiority of the UBV over
the  Str\"omgren system, but to the properties of the BaSeL models.\\
Finally, in order to simultaneously and accurately determine the stellar parameters 
T$_{\rm eff}$, [Fe/H] and log g, we suggest to use the combination of the synthetic BaSeL 
indices B$-$V, U$-$B and b$-$y rather than B$-$V, U$-$B, b$-$y, m$_1$ and c$_1$.  
}

\maketitle

\section{Introduction}
The Basel Stellar Library (BaSeL) is a library of theoretical spectra 
corrected to provide synthetic colours consistent with empirical colour-temperature 
calibrations at all wavelengths from the near-UV to the far-IR (see Cuisinier {\al} 1996 
for the correction procedure, and Lejeune {\al} 1998 and references therein for a complete 
description). 
These model spectra cover a large range of fundamental parameters 
(2000 $\leq$ T$_{\rm eff}$ $\leq$ 50,000 K, $-$5 $\leq$ [Fe/H] $\leq$ 1 
and $-$1.02 $\leq$ log g $\leq$ 5.5) 
and their photometric calibrations are regularly updated (for instance, the  
calibrations in the very low metallicity range is being improved  by comparisons with 
globular cluster populations, Westera et al. 1999) and extended to an even larger set of 
parameters (see Lejeune {\al} 1997, Lejeune {\al} 1998, and Lejeune {\al} 2000). 
\\
The BaSeL library spectra have been calibrated directly for standard dwarf and 
giant sequences at solar abundances and using UBVRIJHKLM broad-band photometry, and 
are hence expected to provide excellent results in these photometric bands.
Since they are based on synthetic spectra, they can in principle be used in many 
other photometric systems taken either individually or simultaneously, and this is 
another major advantage of these models. 
For these reasons, they are currently used in an increasing number of astrophysical 
studies (e.g. Bruzual {\al} 1997, Weiss \& Salaris 1999, Kurth {\al} 1999). \\
Nevertheless, little attention has been paid until now to validate the possibilities 
offered by the BaSeL models in several photometric bands simultaneously.  
The fact that the library spectra have been calibrated using broad-band observations 
(UBVRIJHKLM) does not necessarily imply that similarly reliable results can be expected 
from their application to intermediate- and/or narrow-band photometry data. 
On the contrary, the extent to which the {\it coarse} UBVRIJHKLM calibration 
provides useful results even at higher resolution requires separate investigation. 
In this paper, we investigate the BaSeL models using Johnson and Str\"omgren photometric 
systems simultaneously.
BaSeL Str\"omgren photometry has been previously tested with success against  
stars belonging to double-lined eclipsing binaries for which gravities are accurately 
known (Lastennet {\al} 1999a).  
Here we use a sample of 9 F-type stars for which detailed spectroscopic analysis has  
been performed. 
These stars are potential candidates of the COROT (COnvection and ROTation) space experiment 
main programme (see for instance Catala {\al} 1995, Mi\-chel {\al} 1998, and Baglin {\al} 
1998 for details), and has been observed at the 193cm telescope at Observatoire de 
Haute-Provence (OHP, France) as part of the target selection process.
High signal to noise ratio (S/N $\simeq$ 150) spectra have been obtained using 
{\it Elodie} echelle spectrograph and analysed by comparison with theoretical spectra 
(Ligni\`eres {\al} 1999). 
This set of new spectroscopic data together with both Johnson and Str\"omgren photometry 
from the literature provides a unique test of the predicted results of the BaSeL 
library.
In addition, other calibration methods, namely the Templogg programme (based on Napiwotzki 
{\al} 1993, and K\"unzli {\al} 1997) and the Marsakov \& Shevelev (1995) catalogue, 
provide further comparisons. \\
The paper is organized as follows: Sect. 2 deals with the description of our working 
sample of stars and the extinction issue, Sect. 3 presents the various methods 
used to derive estimates of T$_{\rm eff}$, [Fe/H], and log g, and Sect. 4 is devoted 
to the presentation and the discussion of the results, with particular emphasis put on 
the intrinsic properties of the BaSeL library spectra in Sect. 5. 
Finally, Sect. 6 draws our general conclusions. 

\section{The sample: relevant data and reddening}

\subsection{Basic data for the 9 stars}

The photometric data in the Johnson (B$-$V and U$-$B) and Str\"omgren 
(b$-$y, m$_1$, c$_1$) systems of our working sample of F-type stars are presented 
in Table 1, along with cross-identifications, Hipparcos parallaxes ($\pi$) and 
rotational velocities (v sini is derived from the application of the Least-Squares 
Deconvolution method of Donati {\al} 1997 on the OHP spectra). 

\subsection{Reddening}

Estimates of the extinction are required for photometric calibration methods and in 
particular for the BaSeL and Templogg methods used in this paper. 
Parallaxes listed in Table 1 from the Hipparcos catalogue (ESA, 1997, see also Perryman 
{\al} 1997) show that all the stars are in the close solar neighborhood ($\leq$ 45 pc). 
While reddening is generally expected to be close to zero inside this very local sphere 
(e.g. Welsh {\al} 1991), 
we checked this by applying the extinction model of Vergely (1998) to our stars. 
This model has been constructed from a sample of 4000 stars (in a sphere of 250 pc) 
with Str\"omgren photometry and provides extinction as a function of position and distance
\footnote{However, it is also worth noticing that 
this model neglect small clumps and  circumstellar extinction, which means that 
if there is a small clump with a high density on the line of sight, the model underestimates 
the extinction}. 
The $\sigma_{\pi}$/$\pi$ values ($\leq$3.3\%) reported for the present stars in the Hipparcos 
Catalogue ensure that our extinction estimates cannot be seriously affected by distances 
uncertainties.
Results are shown in Table 1 (col. E(b$-$y)) and confirm that the reddening is very 
close to zero for all stars. Such small values have no influence on the  
inferred photometric results presented in this paper, and will thus be neglected. 
Therefore, all the results obtained in the remainder of this paper will be given without reddening. 

\begin{table*}[htb] 
\caption[]{Cross-identifications (HD and HIP numbers), Hipparcos parallaxes (Perryman et al. 1997), 
extinction (E(b$-$y) from Vergely 1998), rotational velocities v sini 
(km s$^{-1}$), and photometric  data used in the BaSeL and Templogg
determinations for the 9 target stars.} \begin{flushleft}
\begin{center}  
\begin{tabular}{rrrrlrrrrrr}
\hline 
\noalign{\smallskip} 
 ID$^{\dag}$ & HD  &  HIP    & $\pi$ (mas) & E(b$-$y) & v sini & B$-$V & U$-$B & b$-$y & m$_1$  & c$_1$ \\
\noalign{\smallskip}
\hline \noalign{\smallskip}
1 &  43587  &   29860  & 51.76$\pm$0.78 & 0.001 & 2       & 0.61$\pm$0.01  & 0.10$\pm$0.01 &  0.384$\pm$0.01 & 0.187$\pm$0.03 & 0.349$\pm$0.03  \\
2 &  43318  &   29716  & 28.02$\pm$0.76 & 0.002 & 5       & 0.49$\pm$0.01  & 0.00$\pm$0.01 &  0.322$\pm$0.03 & 0.154$\pm$0.03 & 0.446$\pm$0.03  \\
3 &  45067  &   30545  & 30.22$\pm$0.92 & 0.001 & 6       & 0.56$\pm$0.01  & 0.07$\pm$0.01 &  0.361$\pm$0.02 & 0.168$\pm$0.03 & 0.396$\pm$0.03  \\
4 &  49933  &   32851  & 33.45$\pm$0.84 & 0.001 & 10      & 0.39$\pm$0.01  & $-$0.09$\pm$0.02 &  0.270$\pm$0.02 & 0.127$\pm$0.04 & 0.460$\pm$0.03   \\
5 &  49434  &   32617  & 24.95$\pm$0.75 & 0.002 & 79      & 0.295$\pm$0.015 & 0.05$\pm$0.03 &  0.178$\pm$0.01 & 0.178$\pm$0.02 & 0.717$\pm$0.03  \\
6 &  46304  &   31167  & 23.13$\pm$0.76 & 0.002 & 200     & 0.25$\pm$0.01  & 0.06$\pm$0.01 &  0.158$\pm$0.01 & 0.175$\pm$0.02 & 0.816$\pm$0.04  \\
7 &  162917 &   87558  & 31.87$\pm$0.77 & 0.004 & 25      & 0.42$\pm$0.01  & $-$0.03$\pm$0.02 &  0.280$\pm$0.02 & 0.166$\pm$0.06 & 0.458$\pm$0.01  \\
8 &  171834 &   91237  & 31.53$\pm$0.75 & 0.004 & 64      & 0.37$\pm$0.01  & $-$0.04$\pm$0.01 &  0.254$\pm$0.04 & 0.145$\pm$0.01 & 0.560$\pm$0.04  \\
9 &  164259 &   88175  & 43.11$\pm$0.75 & 0.003 & 76      & 0.38$\pm$0.01  & $-$0.01$\pm$0.01 &  0.253$\pm$0.01 & 0.153$\pm$0.01 & 0.560$\pm$0.04    \\
\noalign{\smallskip}\hline        
\end{tabular}
\end{center}
$^{\dag}$ Arbitrary running number. \\
\end{flushleft}
\end{table*}

\section{Description of the different methods}
In this section, we present the various methods used to derive T$_{\rm eff}$, [Fe/H], 
and log g. Apart from the BaSeL models, a detailed spectroscopic analysis 
and two photometric calibrations have been considered. 

\subsection{Photometric analysis}

\subsubsection{The BaSeL models} 

For the present work, more than 50700 models have been computed by interpolation by two of us 
(Lastennet and Lejeune), each model giving synthetic photometry in the Johnson 
and Str\"omgren systems for a set of (T$_{\rm eff}$, [Fe/H], log g). 
In order to fit the observed colours of the target stars, we have computed a fine 
grid in the (T$_{\rm eff}$, [Fe/H], log g) parameter space. 
The grid explored is defined in this way: 5000 $\leq$ T$_{\rm eff}$ $\leq$ 8000K in 20K steps, 
$-$1 $\leq$ [Fe/H] $\leq$ 0.5 in 0.1 steps, and 3 $\leq$ log g $\leq$ 5 in 0.1 steps. 
These parameter ranges are reasonable matches to the expected properties of F dwarfs, 
for which the BaSeL library includes the Kurucz atmospheric models (Mixing Length
Theory of convection with $l/ H_{\rm P} = 1.25$, and with the overshooting parameter 
equal to 1).

In order to derive simultaneously the effective temperature (T$_{\rm eff}$), the \met 
([Fe/H]), and the surface gravity (log g) of each star, we minimize the $\chi^2$-functional 
defined as: 

\beqa
\chi^2 (T_{\rm eff}, [Fe/H], \log g) & = & 
\sum_{i=1}^{5} \left[ \left(\frac{\rm col(i)_{\rm mod} - col(i)}
{\sigma(\rm col(i))}\right)^2 \right]  \nonumber \\
\eeqa
where col(i) and $\sigma$(col(i)) are the observed values (B$-$V,
U$-$B, b$-$y, m$_1$, and c$_1$) and their error bars, as reported
in Table 1, and col(i)$_{\rm mod}$  are obtained from the synthetic
computations of the BaSeL models. 
A similar method has already been developed
and used by Lastennet {\al} (1996, 1999b)  for CMD diagrams. \\
With 5 observational data (B$-$V, U$-$B, b$-$y, m$_1$ and c$_1$ for each star) and 
3 free parameters (T$_{\rm eff}$, log g and [Fe/H]), we expect to 
find a $\chi^2$-distribution with 2 degrees of freedom.
Since the $\chi^2$-value is an estimation of the quality of the fit, with 2 degrees of 
freedom (DOF), a $\chi^2$-value smaller than 6 is a good fit, because it means that 
the probability P that this $\chi^2$$_{DOF=2}$ is smaller than 6 is about 95\%, i.e. 
P($\chi^2$$_{DOF=2}$) $<$ 6 $\simeq$ 95 \%. This criterion will change according to the 
DOF, but basically, small values are signatures of good fits. 
In the remainder of this paper, the $\chi^2$-values 
generally satisfy this criterion  being most of the time much better,  
hence by default they will not be discussed when they are good. 
A $\chi^2$-grid is formed in the (T$_{\rm eff}$, log g, [Fe/H]) parameter space. 
Once the central minimum value $\chi^{2}_{\rm min}$ is found, we compute the surfaces  
corresponding to 1$\sigma$, 2$\sigma$, and 3$\sigma$ confidence levels. 
For clarity, the intersection of these confidence surfaces with the appropriate 
plane will be displayed (e.g. T$_{\rm eff}$-[Fe/H] in Fig.1, T$_{\rm eff}$-log g in Fig. 2). 
\\  
  
\subsubsection{The "Templogg" method}

We have also run the "Templogg" program which is designed to determine 
effective temperature and log g from either Str\"omgren or Geneva photometry. 
For Str\"omgren photometry it uses a Fortran program written by E. Fresno
which relies  upon the grids of Moon \& Dworetsky (1985) in the T$_{\rm
eff}$-log g parameter  space relevant to this paper, with the improvements by
Napiwotzki {\al} (1993). 
For Geneva photometry it uses a Fortran program
written by M. Kunzli  (see North {\al} 1994). 
The program chooses among eight different regions in the HR diagram for 
selecting the best calibration within the Str\"omgren system and three 
different regions for the Geneva system. 
The results from this method are gathered in Table 2. 

\subsubsection{The catalogue of Marsakov \& Shevelev (1995)}

Marsakov \& Shevelev (1995) (hereafter [MS95]) have computed effective temperatures 
and surface gravities using Moon's (1985) method, which is also based on the 
interpolation of the grids presented in Moon \& Dworetsky (1985). 
According to Moon (1985), the standard deviations of the derived 
parameters are T$_{\rm eff}$$=$$\pm$ 100 K and log g $=$ $\pm$0.06.
The metallicities of Marsakov \& Shevelev (1995) are obtained with the 
equation of Carlberg {\al} (1985): 

\beqa
[Fe/H] = 0.16 - 0.66 \times \Delta\beta - 
\left[12.3 -38 \times \Delta\beta\right] \times \delta m_1, \nonumber\\
\eeqa

\noindent where $\Delta$$\beta$$=$2.72 $-$ $\beta$, and the colour excess $\delta$m$_1$ 
and the Str\"omgren $\beta$ index are determined from (b$-$y). 
All the [MS95] results relevant for our sample are given in Tab. 2. 

\subsection{Spectroscopic analysis}

The fundamental stellar parameters have been derived from a detailed analysis 
of spectra with high signal to noise ratio (S/N $\simeq$ 150) obtained at 
OHP with the {\it Elodie} echelle spectrograph (spectra ranging from 3906 \AA \, to 
6811 \AA \, at a resolution of $\lambda/\Delta \lambda$ $=$ 42000). 
After reduction, the observed spectra were compared with theoretical ones constructed 
from a combination of Kurucz atmospheric models (ATLAS9 - Mixing Length
Theory of convection with $l/ H_{\rm P} = 0.5$ and without overshooting. 
This choice of parameters different than those used by Kurucz is fully
justified in van't Veer-Menneret \& M\'egessier (1996)), 
the VALD-2 atomic database (Kupka {\al} 1999), 
and the SYNTH radiative transfer  codes (Piskunov 1992) and BALMER9 (Kurucz
1993).  In addition, the Least-Squares Deconvolution  method (Donati {\al}
1997) provided accurate determination of the projected rotational  velocities
listed in Tab. 1.  Details about these determinations are given in the two
next subsections and the results  are summarized in Tables 1 and 2. Note that
the level of accuracy of these results, initially  designed to select COROT
targets, is sufficient for the purpose of the present study.

\subsubsection{Determination of T$_{\rm eff}$ from the H$\alpha$ line}

The effective temperature can be determined by taking advantage of the sensitivity of 
H$\alpha$ line wings. Most importantly, detailed studies (e.g. van 't Veer-Menneret \& 
M\'egessier 1996), have shown that the H$\alpha$ line is independent of the surface gravity 
(for non-supergiant stars) for effective temperatures ranging from 5000 K to $\sim$8500 K, 
and depends only slightly on the metallicity. T$_{\rm eff}$ is therefore obtained for each 
star of the sample by fitting the observed H$\alpha$ line with synthetic spectra computed 
from a grid of solar metallicity atmospheric models separated by 250K. 

\subsubsection{Determination of the surface gravity and metallicity}

Within the temperature range that we found, we noticed that Fe I absorption lines
depend only on temperature and metallicity, being practically independent of the surface
gravity, while Fe II lines are sensitive to the temperature, metallicity and gravity.
Consequently, the temperature being known from the H$\alpha$ line, the metallicity
along with the microturbulence velocity are determined first by fitting a set of weak
and strong Fe I lines.
Then, the gravity is obtained by fitting Fe II lines. 
The spectral region near 6130 \AA \, proved to be suitable for this analysis. 
However, the line broadening induced by rotation tends to mix neighbouring lines and 
prevent the analysis of individual Fe lines for high values of v sini 
(HD 49434, HD 46304, HD 171834 and HD 164259).

\subsection{Other determinations in the literature} 

To be as complete as possible, we looked for other determinations available in the literature 
and the SIMBAD database. 
One of the most comprehensive sources for our purpose is the fifth Edition of the catalogue of 
Cayrel de Strobel {\al} (1997), which includes [Fe/H] determinations and atmospheric parameters 
($T_{\rm eff}$, log g) obtained from high-resolution spectroscopic observations and detailed 
analyses, most of them carried out with the help of model atmospheres. 
However, the stars of our sample are not included in this catalogue. Since the catalogue 
comprises the literature (700 bibliographical references) up to December 1995, 
we only looked for more recent references. 
To the best of our knowledge, the catalogue of metallicities of Zakhozhaj 
\& Shaparenko (1996) (hereafter [ZS96]) is the only one which contains useful information 
for our purpose. 
These metallicities are obtained from photometric UBV data and are available for 
two stars of our sample: HD 43587 and HD 164259 (see Table 2).

\begin{table*}[htb]
\caption[]{Comparison of fundamental stellar parameters determinations. The {\it Templogg} method 
uses Str\"omgren and Geneva photometric data,  Marsakov \& Shevelev 1995 ({\it [MS95]}) used 
Str\"omgren data, and the {\it BaSeL} method uses Johnson and Str\"omgren data. Results from 
our spectroscopic analysis ({\it Spectro.}), and Zakhozhaj \& Shaparenko 1996 {\it [ZS96]} are also 
given.}
\begin{flushleft}
\begin{center}  
\begin{tabular}{|l|l|ccccccccc|}
\hline\noalign{\smallskip}
ID$^{\dag}$    &            & 1     & 2      & 3      & 4      & 5      & 6      & 7      & 8      &    9  \\
HD             &            & 43587 & 43318  & 45067  & 49933  & 49434  & 46304  & 162917 & 171834 & 164259 \\
\noalign{\smallskip}\hline \noalign{\smallskip} 
               & Method     &  &  &  &  &  &  &  &  &  \\
\noalign{\smallskip}\hline \noalign{\smallskip}
T$_{\rm eff}$  & Templogg    & 6009 & 6420 & 5982 & 6535 & 7321 & 7379 & 6587 & 6714 & 6789 \\
               & [MS95]      & 5952 & 6280 & 6066 & 6625 &      &      & 6629 & 6739 & 6730 \\
               & BaSeL$^{a}$ & 5740 & 6040 & 6000 & 
               6420 & 7120 & 7200 & 6400 & 6580 & 6580 \\
               & BaSeL$^{c}$ & 5720 & 6320 & 5940 & 
               6600 & 7240 & 7240 & 6660 & 6700 & 6820 \\ 
               & Spectro.$^{\ddag}$  & 6000 & 6250 & 6000 & 6500 & 7250 &
7250 & 6500 & 6750 & 6750 \\  \noalign{\smallskip}\hline \noalign{\smallskip}
log g          & Templogg    & 4.32 & 4.20 & 4.16 & 4.25 & 4.16 & 3.93 & 4.32 & 4.02 & 4.11 \\
               & [MS95]      & 4.11 & 4.05 & 4.02 & 4.46 &      &      & 4.49 & 4.10 & 4.10 \\
               & BaSeL$^{a}$ & 3.4  & 3.1 & 3.9 & 3.5 & 3.5 &
               3.3 & 3.6 & 3.4 & 3.3 \\               
               & BaSeL$^{b}$ & bad fit  & 4.5 & 3.8 & 3.9 & 4.0 &
               3.4 & 4.0 & 4.1 & 4.0 \\              
               & BaSeL$^{c}$ & 4.3  & 4.5 & 3.8 & 4.3 & 4.0 &
               3.4 & 4.5 & 3.9 & 4.2 \\                                          
               & Spectro.$^{\ddag}$ & 4.5 & 4.0 & 4.0 & 4.0 & & & 4.0 &  & \\ 
\noalign{\smallskip}\hline \noalign{\smallskip}
[Fe/H]         & Templogg    & $-$0.13 & $-$0.18 & $-$0.15 & $-$0.48 & 
	     $-$0.03 & $-$0.09 & $+$0.03 & $-$0.20 & $-$0.11 \\
               & [MS95]      & $-$0.15 & $-$0.18 & $-$0.17 & $-$0.35 &   &   & $+$0.08 & $-$0.15 & $-$0.05 \\               
               & BaSeL$^{a}$ & $-$0.2 & $-$0.5 & $+$0.0 & $-$1.0 & $-$0.5 &
               $-$0.9 & $-$0.5 & $-$0.8 & $-$0.6 \\              
               & BaSeL$^{b}$ & bad fit & $-$0.1 & $+$0.0 & $-$0.8 & $-$0.1 &
               $-$0.8 & $-$0.3 & $-$0.4 & $-$0.2 \\              
               & BaSeL$^{c}$ & $-$0.2 & $+$0.0 & $-$0.1 & $-$0.6 & $-$0.1 &
               $-$0.8 & $+$0.0 & $-$0.5 & $+$0.0 \\                                                       
               & Spectro.$^{\ddag}$  & $-$0.1 & $-$0.3 & $-$0.1 & $-$0.5 & &
& $-$0.2 &  &  \\                 & [ZS96]      & $+$0.03 &   &   &   &   &  
&   &   & $-$0.03 \\ \noalign{\smallskip}\hline
\end{tabular}
\end{center}
\small
$^{\dag}$ Running number as in Tab. 1. \\
$^{a}$ From UBV Johnson (B$-$V and U$-$B) and Str\"omgren (b$-$y, m$_1$ and c$_1$) data 
(cf. Fig. \ref{f:tefffeh_all} or Fig. \ref{f:tefflogg_all}).\\
$^{b}$ From UBV Johnson data (Fig. \ref{f:loggfeh_J}). \\
$^{c}$ From the combination B$-$V, U$-$B and b$-$y (Fig. \ref{f:final_tefffeh} and \ref{f:final_tefflogg}).\\ 
$^{\ddag}$ Estimated error: $\Delta$T$_{\rm eff}$$\simeq$$\pm$250 K, 
$\Delta$log g$\simeq$$\pm$0.5, $\Delta$[Fe/H]$\simeq$$\pm$0.2. \\
\end{flushleft}
\end{table*}
\vspace{-0.5cm}
\normalsize

\section{Multi-colour exploration of the BaSeL models: comparisons and discussion}

In this section, we will discuss the results obtained from the exploration of the 
BaSeL model properties in the Johnson and Str\"omgren photometric systems (the best solutions  
from the BaSeL models and from the other methods are summarized in Table 2). 
First we discuss the results in the T$_{\rm eff}$, log g, [Fe/H] parameter space 
using all the photometric indices. 
Discrepancies between synthetic and spectroscopic results led us to consider solutions derived 
from Johnson and Str\"omgren photometry separately, by fixing (in the second part of this discussion) 
T$_{\rm eff}$ at its spectroscopic values and investigating the BaSeL properties in log g-[Fe/H] 
diagrams. 
Finally, we present the most reliable combination of synthetic photometry indices
to get T$_{\rm eff}$, log g and [Fe/H] simultaneously.  

\subsection{Simultaneous T$_{\rm eff}$, log g, [Fe/H] determinations using all the photometric indices}

\begin{figure*}[htb]
\centerline{\psfig{file=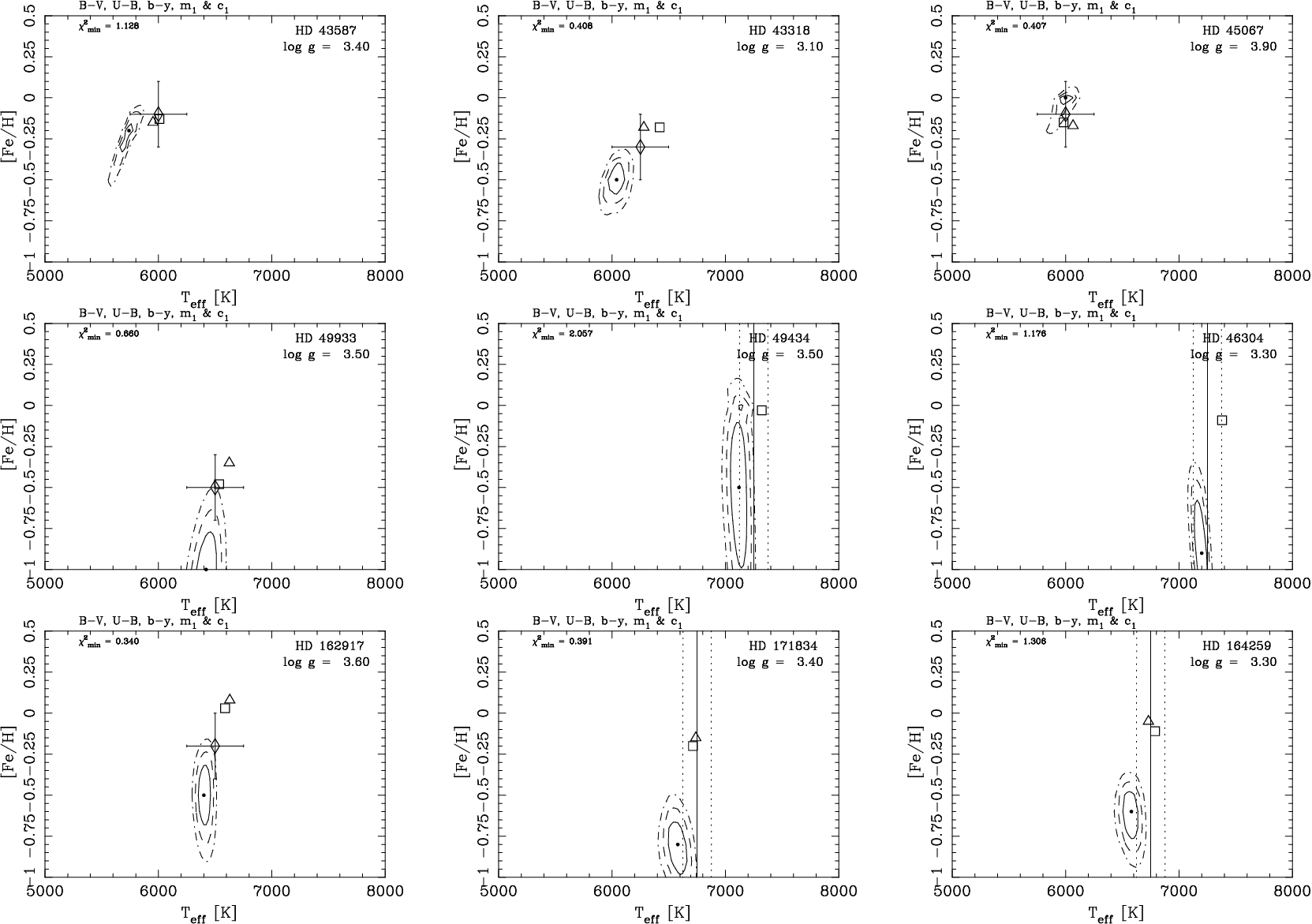,width=18.5truecm,height=19.6truecm,angle=-90.}}
\caption{
Simultaneous (T$_{\rm eff}$, [Fe/H]) results for the 9 potential targets 
of the COROT central seismology programme. The solutions from the BaSeL models 
({\it 1-, 2- and 3-$\sigma$ confidence level contours}) are obtained in order to 
fit simultaneously the 5 following observed photometric values : (B$-$V), (U$-$B), 
(b$-$y), m$_1$ and c$_1$.
For each star, 
the contour solutions are displayed in a log g $=$ constant plane, corresponding 
to the best simultaneous (T$_{\rm eff}$, [Fe/H], log g) solutions derived from 
the BaSeL models (the grid explored is: 5000 $\leq$ T$_{\rm eff}$ $\leq$ 8000 K 
in 20K steps, $-$1 $\leq$ [Fe/H] $\leq$ 0.5 in 0.1 steps, and 3 $\leq$ log g $\leq$ 5  
in 0.1 steps). 
An estimation of the quality of the best fit ($\chi^2$-value) is also quoted in each panel. 
The results projected in the T$_{\rm eff}$-[Fe/H] planes from the
spectroscopic analysis ({\it diamond with error bars}, or {\it solid plus two dotted lines}) 
as well as from the "Templogg" programme ({\it square}), 
and Marsakov \& Shevelev (1995) ({\it triangle}) are also shown for comparison.
}
\label{f:tefffeh_all}
\end{figure*}

\begin{figure*}[htb]
\centerline{\psfig{file=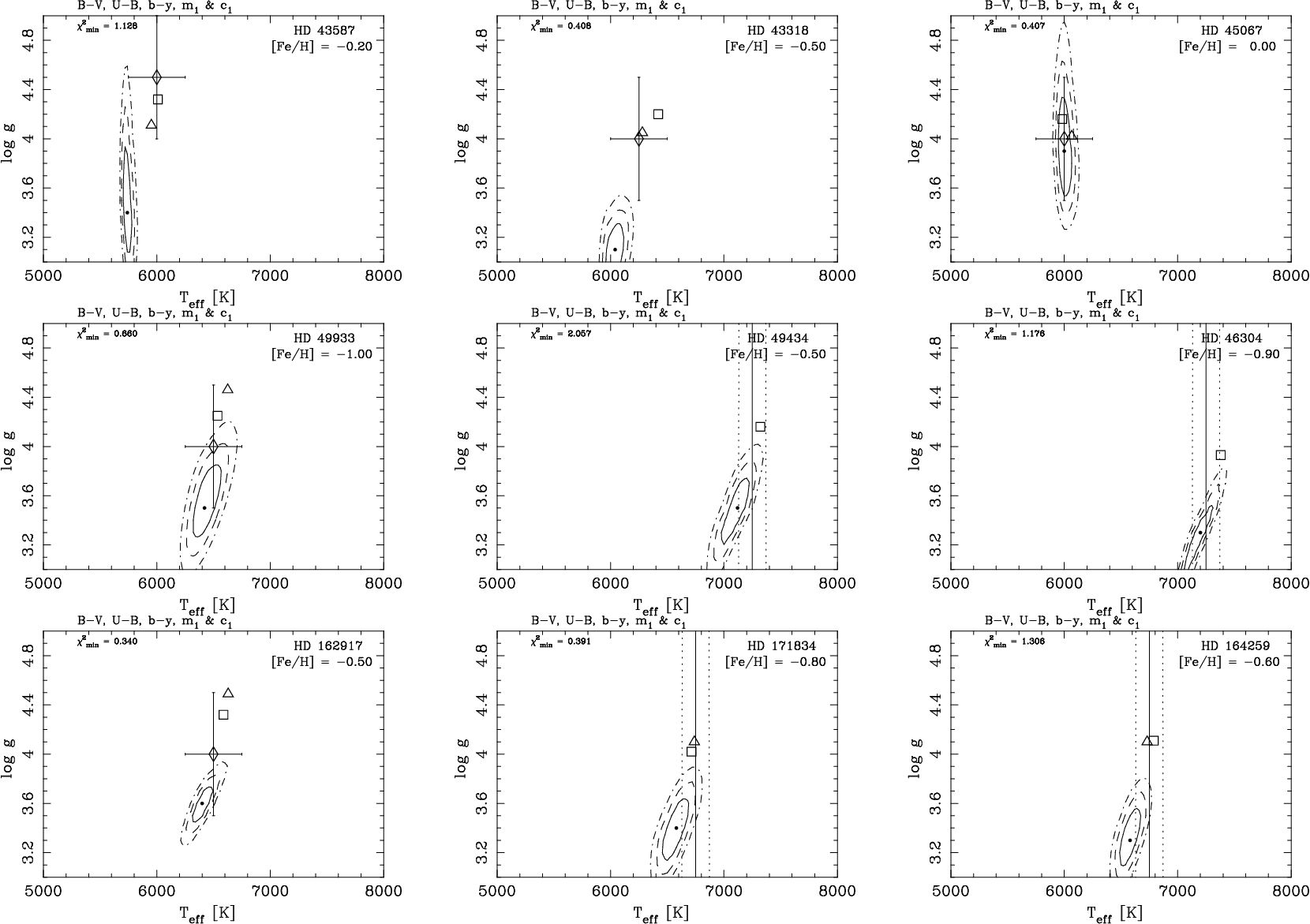,width=18.5truecm,height=19.6truecm,angle=-90.}}
\caption{As Fig. \ref{f:tefffeh_all}, but this time in a (T$_{\rm eff}$, log g) 
diagram. 
For each star, the contour solutions are displayed in a [Fe/H] $=$ constant plane, 
corresponding to the best simultaneous (T$_{\rm eff}$, [Fe/H], log g) solutions derived 
from the BaSeL models. 
}
\label{f:tefflogg_all}
\end{figure*}

The results obtained with all the methods described in Sec. 3 are shown in T$_{\rm eff}$-[Fe/H] 
diagrams (Fig. \ref{f:tefffeh_all}) and T$_{\rm eff}$-log g diagrams (Fig. \ref{f:tefflogg_all}) 
for the 9 stars of our sample. 
The solutions from the BaSeL models (i.e. the best $\chi^2$-solution plus the 1-, 2- and 3-$\sigma$ 
confidence level contours) are 
obtained in order to fit simultaneously the 5 available observed photometric values of Tab. 1: 
(B$-$V), (U$-$B), (b$-$y), m$_1$, and c$_1$.
For each star, the contour solutions are displayed in a log g $=$ constant plane 
(Fig. \ref{f:tefffeh_all}) or an [Fe/H] $=$ constant plane (Fig. \ref{f:tefflogg_all}), 
corresponding to the best simultaneous (T$_{\rm eff}$, [Fe/H], log g) solutions derived from 
the BaSeL models.
When available, the results from the spectroscopic analysis (diamond with error bars) as 
well as from the "Templogg" programme (square), and Marsakov \& Shevelev (1995) (triangle) 
are projected in these diagrams for comparison. \\
As far as the effective temperature is concerned, the overall agreement is good between all 
the methods.  
However, with comparison to spectroscopy, the BaSeL models give systematically lower values: 
$<$$\Delta$T$_{\rm eff}$$>$$=$ $<$T$_{\rm eff}$(spectro) $-$ T$_{\rm eff}$(BaSeL)$>$ 
$=$ 130K on average (which is comparable to the result of Cuisinier {\al} 1994). 
On the other hand, the Templogg method gives systematically larger values (by $\sim$ 60K on 
average). Results given by either the [MS95] catalogue or the Templogg method are essentially 
identical.\\
There is no such overall agreement for the surface gravity and the metallicity, 
and this disagreement is neither correlated with increasing temperature nor with increasing v sini. 
For instance in Fig. \ref{f:tefflogg_all}, while the systematic disagreement in gravity appears 
for stars with the hottest spectroscopic temperatures and high rotation speeds (HD 49434, 
HD 46304, HD 171834 and HD 164259), the disagreement is even worse for the cooler and slowly 
rotating star HD 43318.  
Thus, except in one case (HD 45067) where the match is perfect for both log and [Fe/H], 
BaSeL-derived metallicities and gravities are systematically lower than the values obtained from 
the other methods. 
In particular, the best-$\chi^2$ values of log g inferred from the synthetic BaSeL colours  
range between 3.1 and 3.9, indicating possible evolved/sub-giant stars, which is ruled out 
by spectroscopic determinations (listed in Tab. 2). 
Even if inspection of the confidence contours might temper this conclusion by pointing out 
that the BaSeL results are sometimes compatible to 3-$\sigma$, the systematic behaviour 
remains, at least for the small sample studied, and requires further analysis. 

\subsection{Simultaneous [Fe/H]-log g determinations using various combinations of photometric indices} 
\begin{figure*}[htb]
\centerline{\psfig{file=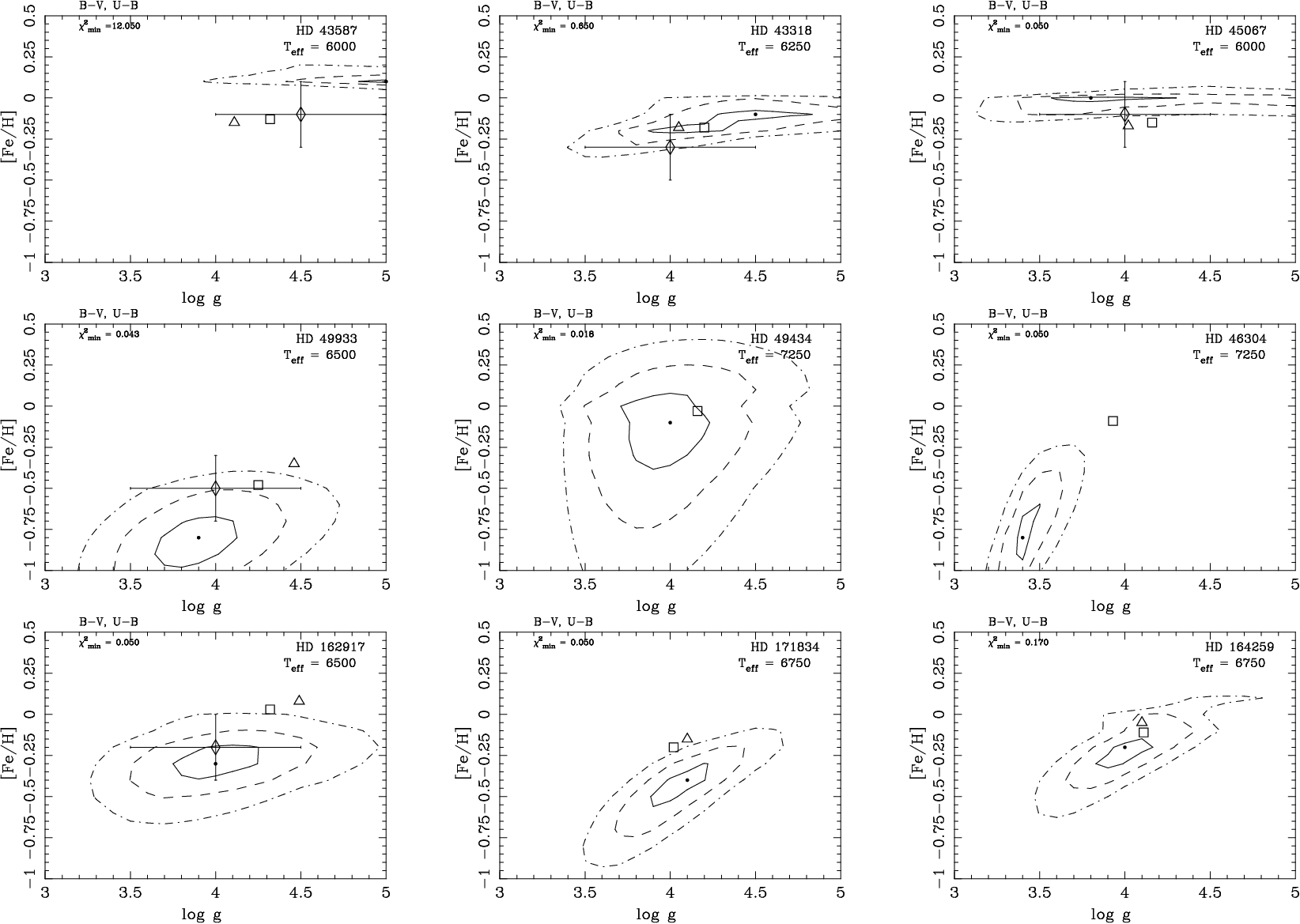,width=18.5truecm,height=19.6truecm,angle=-90.}}
\caption{
Simultaneous (log g, [Fe/H]) results for the 9 potential targets 
of the COROT central seismology programme. The solutions from the BaSeL models 
({\it 1-, 2- and 3-$\sigma$ confidence level contours}) are obtained in order to 
fit simultaneously the 2 following observed photometric values : (B$-$V) and (U$-$B). 
For each star, the contour solutions are displayed in a T$_{\rm eff}$ $=$ constant plane, 
corresponding to the best T$_{\rm eff}$ spectroscopic estimates. The $\chi^2$$-$value is 
an estimation of the quality of the fit (a $\chi^2$-value close to 0 is a good fit). Only 
one bad fit is obtained: $\chi^2$$\simeq$12 for HD 43587 ({\it upper left panel}).
The results projected in the log g-[Fe/H] plane from the spectroscopic analysis 
({\it diamond with error bars}) as well as 
from the "Templogg" programme ({\it square}), and Marsakov \& Shevelev (1995) ({\it triangle}) 
are also shown for comparison.
}
\label{f:loggfeh_J}
\end{figure*}

\begin{figure*}[htb]
\centerline{\psfig{file=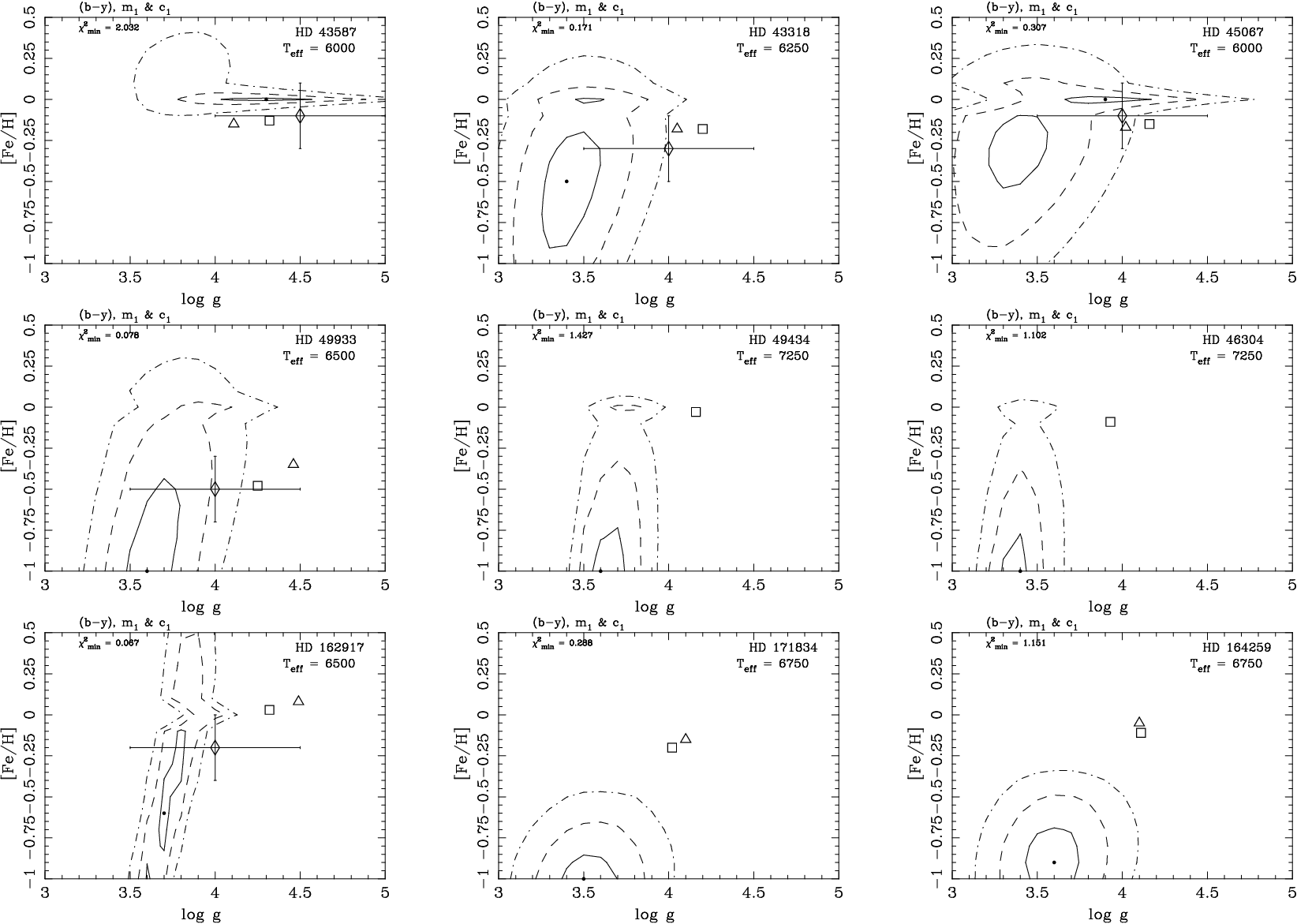,width=18.5truecm,height=19.6truecm,angle=-90.}}
               \caption{
Same as Fig.~\ref{f:loggfeh_J}, but the solutions from the BaSeL models 
are obtained in order to fit simultaneously the 3 following observed photometric 
values : (b$-$y), m$_1$, and c$_1$. 
}
\label{f:loggfeh_S}
\end{figure*}

Since it appears that the BaSeL models have difficulties reproducing simultaneously 
the 3 parameters with all the photometric information (Str\"omgren plus Johnson), 
we checked if this remains true using either Johnson or Str\"om\-gren photometry 
separately, in order to find out which photometric strategy would improve the situation. 
As only two Johnson indices are available in the data (Table 1), namely B$-$V and U$-$B, 
the three fundamental parameters cannot be derived simultaneously using Johnson 
photometry alone. 
Therefore, we choose to fix the effective temperature derived from spectroscopy, and 
results for Johnson (Fig. \ref{f:loggfeh_J}) and Str\"omgren (Fig. \ref{f:loggfeh_S}) 
are compared in log g-[Fe/H] diagrams. 
Note that the m$_1$ and c$_1$ indices are optimized linear combinations of colour indices
designed for the purposes of measuring metallicity and surface 
gravity, respectively, whereas UBV was not designed at all on such physical considerations. 
Therefore, the m$_1$ and c$_1$ indices should {\it a priori} give the best determinations 
of [Fe/H] and log g. 
\\
The effective temperature being fixed to its spectroscopic value, 
the \met and the surface gravity of each star are determined simultaneously 
by minimizing the following $\chi^2$-functional:

\beqa
\chi^2 ([Fe/H], \log g) & = & 
\sum_{i=1}^{2} \left[ \left(\frac{\rm col(i)_{\rm mod} - col(i)}
{\sigma(\rm col(i))}\right)^2 \right]  \nonumber \\
\eeqa
where col(i) and $\sigma$(col(i)) are the observed values and their error
bars respectively, and col(i)$_{\rm mod}$ are obtained from the synthetic
computations of the BaSeL models.

Figures \ref{f:loggfeh_J} and \ref{f:loggfeh_S} display the log g-[Fe/H] solutions  
for Johnson and Str\"omgren, respectively.  
The impressively good results of Fig. \ref{f:loggfeh_J} show that for the spectroscopic 
temperatures, the BaSeL library spectra provide log g values which deviate from the 
spectroscopic values neither by a significant amount, nor in a systematic manner. \\
A similar, only slightly weaker agreement holds for the metallicities: they come out 
within less than $\sim$ 0.2 dex from the spectroscopic results, which is excellent in 
view of the intrinsic accuracy ($\sim$ 0.2 dex) that can be obtained 
from empirical UBV data and calibrations (e.g. Buser \& Kurucz 1992). 

\begin{figure*}[htb]
\centerline{\psfig{file=MS9834f5.eps,width=18.5truecm,height=19.6truecm,angle=-90.}}
               \caption{
Same as Fig.~\ref{f:loggfeh_J}, but the solutions from the BaSeL models 
are obtained in order to fit simultaneously the 3 following observed photometric 
values : (B$-$V), (U$-$B), (b$-$y). 
}
\label{f:loggfeh_BVUBby}
\end{figure*}

\begin{figure}[htb]
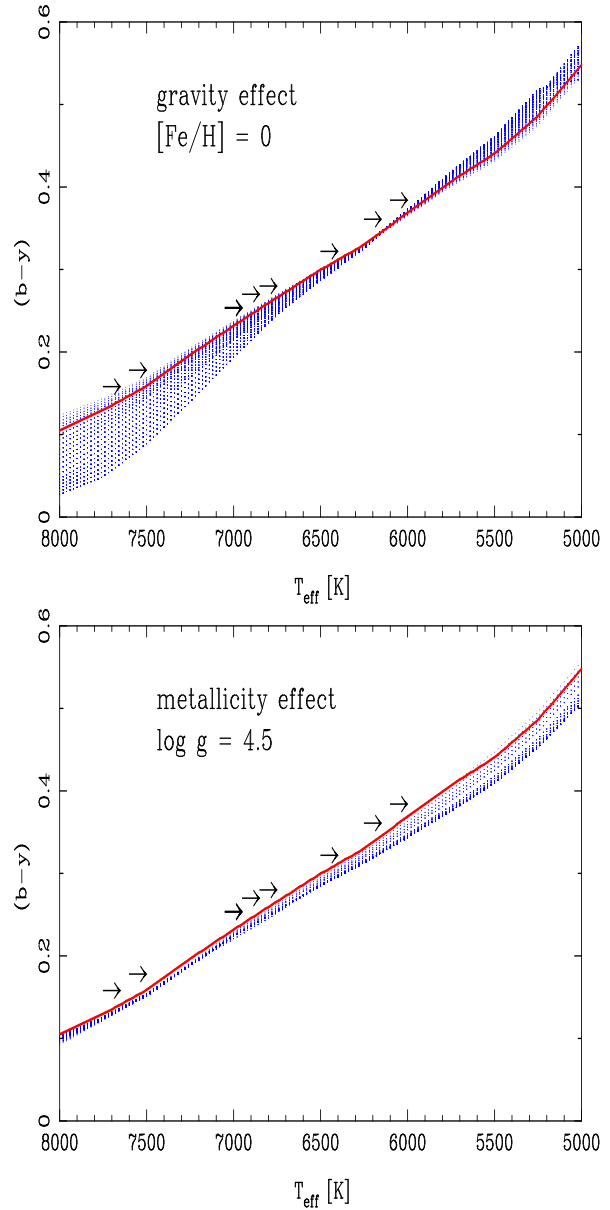

 \centerline{\psfig{file=MS9834f6a.eps,width=7.80truecm,height=8.truecm,angle=-90.}}
 \centerline{\psfig{file=MS9834f6b.eps,width=7.80truecm,height=8.truecm,angle=-90.}}
\caption{T$_{\rm eff}$-(b$-$y) BaSeL relationship for the effective temperature range 
(5000-8000 K) relevant for this work. 
On the top panel the effect of gravity (dashed area) is shown for [Fe/H] $=$ 0 and log g values 
between 3 and 5, and on the lower panel the effect of metallicity (dashed area) is shown for 
log g $=$ 4.5 and [Fe/H] between $-$1. and 0.5. 
For comparison a [Fe/H] $=$ 0 and log g $=$ 4.5 model (bold solid line), and 
the (b$-$y) values (arrows) of the sample stars are shown on both panels. 
}
\label{f:effects} 
\end{figure}

Direct comparison of Fig.\ref{f:loggfeh_J} with Fig. \ref{f:loggfeh_S} - which uses the 
information on all the Str\"omgren indices - clearly shows that the results from 
Johnson and Str\"omgren are not compatible. 
In most cases, even the 1-$\sigma$ contours are partly outside the log g-[Fe/H] diagrams 
in Fig. \ref{f:loggfeh_S}, suggesting that the best solutions inferred from the 3 Str\"omgren 
synthetic indices are perhaps even further away from the expected values. 
We performed various tests to identify which synthetic Str\"omgren index (indices) is (are) 
responsible for the disagreement with the Johnson solutions. 
We add separately each Str\"omgren index to the UBV data, in order 
to highlight the influence of each of them on the contours derived from the Johnson data. 
The indices m$_1$ and c$_1$ are clearly found to modify the Johnson contours, in the sense 
of degrading the results. 
On the other hand, b$-$y does not modify the UBV confidence contours. 
This last point is illustrated by comparing Fig. \ref{f:loggfeh_J} (B$-$V, U$-$B) with 
Fig. \ref{f:loggfeh_BVUBby} (B$-$V, U$-$B and b$-$y). 
We observe that neither the contours nor the $\chi_{\rm min}$ values are modified by 
the addition of the b$-$y index to the B$-$V and U$-$B indices.  

\begin{table}[htb] 
\caption[]{Comparison of T$_{\rm eff}$(spectro) vs. T$_{\rm eff}$(BaSeL) using 
b$-$y only. The two values from the BaSeL models 
correspond to the maximum width due to the effect of surface gravity (at [Fe/H]$=$0) 
and the effect of metallicity (at log g$=$4.5) as shown on Fig.~\ref{f:effects}.}
\begin{flushleft}
\begin{center}  
\begin{tabular}{rrccc}
\hline 
\noalign{\smallskip}
 ID$^{\dag}$ & HD  &  T$_{\rm eff}$$^{\ddag}$ &  \multicolumn{2}{c}{ T$_{\rm eff}$(BaSeL) } \\
             &     &    (spectro)            &  3$\leq$log g$\leq$5  &  $-$1$\leq$[Fe/H]$\leq$0.5  \\
\noalign{\smallskip}
\hline \noalign{\smallskip}
1 &   43587  & 6000  & 5860-5940  & 5680-5910  \\
2 &   43318  & 6250  & 6260-6315  & 6160-6320  \\
3 &   45067  & 6000  & 6020-6070  & 5850-6060  \\
4 &   49933  & 6500  & 6590-6750  & 6610-6730  \\
5 &   49434  & 7250  & 7055-7440  & 7305-7380  \\
6 &   46304  & 7250  & 7150-7610  & 7440-7510  \\
7 &   162917 & 6500  & 6525-6665  & 6525-6650  \\
8 &   171834 & 6750  & 6685-6875  & 6730-6850  \\
9 &   164259 & 6750  & 6685-6875  & 6730-6850  \\
\noalign{\smallskip}\hline
\end{tabular}
\end{center}
$^{\dag}$ Running number as in Tab. 1 \\
$^{\ddag}$ Estimated error: $\Delta$T$_{\rm eff}$$\simeq$$\pm$250 K \\
\end{flushleft}
\end{table}

The reason of this remarkable property is twofold: 
first, the b$-$y index depends only slightly on log g and [Fe/H] in the parameter range 
considered; and second, the BaSeL T$_{\rm eff}$ associated with b$-$y is fully compatible 
with the spectroscopic T$_{\rm eff}$ for the 9 stars.  
The first point is illustrated in Fig. \ref{f:effects} in the effective temperature range 
5000-8000 K: the upper panel shows the effect of gravity (log g between 3 and 5) for constant 
metallicity ([Fe/H] $=$ 0), and the lower one the effect of metallicity ([Fe/H] between $-$1 
and 0.5) for constant gravity (log g $=$ 4.5). 
It appears clearly from these two plots that the T$_{\rm eff}$$=$f(b$-$y) relationship is little 
influenced by surface gravity between $\sim$ 5800 and 6700 K, and by metallicity above 6500 K. 
The second point is proved in Table 3, where T$_{\rm eff}$ derived from the
BaSeL calibrations  shown in Fig. \ref{f:effects} are in agreement with the
spectroscopic determinations.   This underlines the reliability of temperature
determinations based on the synthetic  b$-$y index derived from the BaSeL
models. \\ In conclusion, we find that synthetic Str\"omgren m$_1$ and c$_1$
account for  the difference between the Johnson and Str\"omgren solutions.
We also find that B$-$V, U$-$B are the best BaSeL colours to derive log g and [Fe/H] when 
T$_{\rm eff}$ is fixed. 
Since b$-$y gives good T$_{\rm eff}$ estimates, this hence suggests that the B$-$V, U$-$B, b$-$y 
combination, i.e. a combination of Johnson {\it and} Str\"omgren photometric indices, 
should be able to produce reliable simultaneous estimates of the 3 fundamental parameters. 
This suggestion is developed in the next part. 

\subsection{Best synthetic colour combination}

\begin{figure*}[htb]
\centerline{\psfig{file=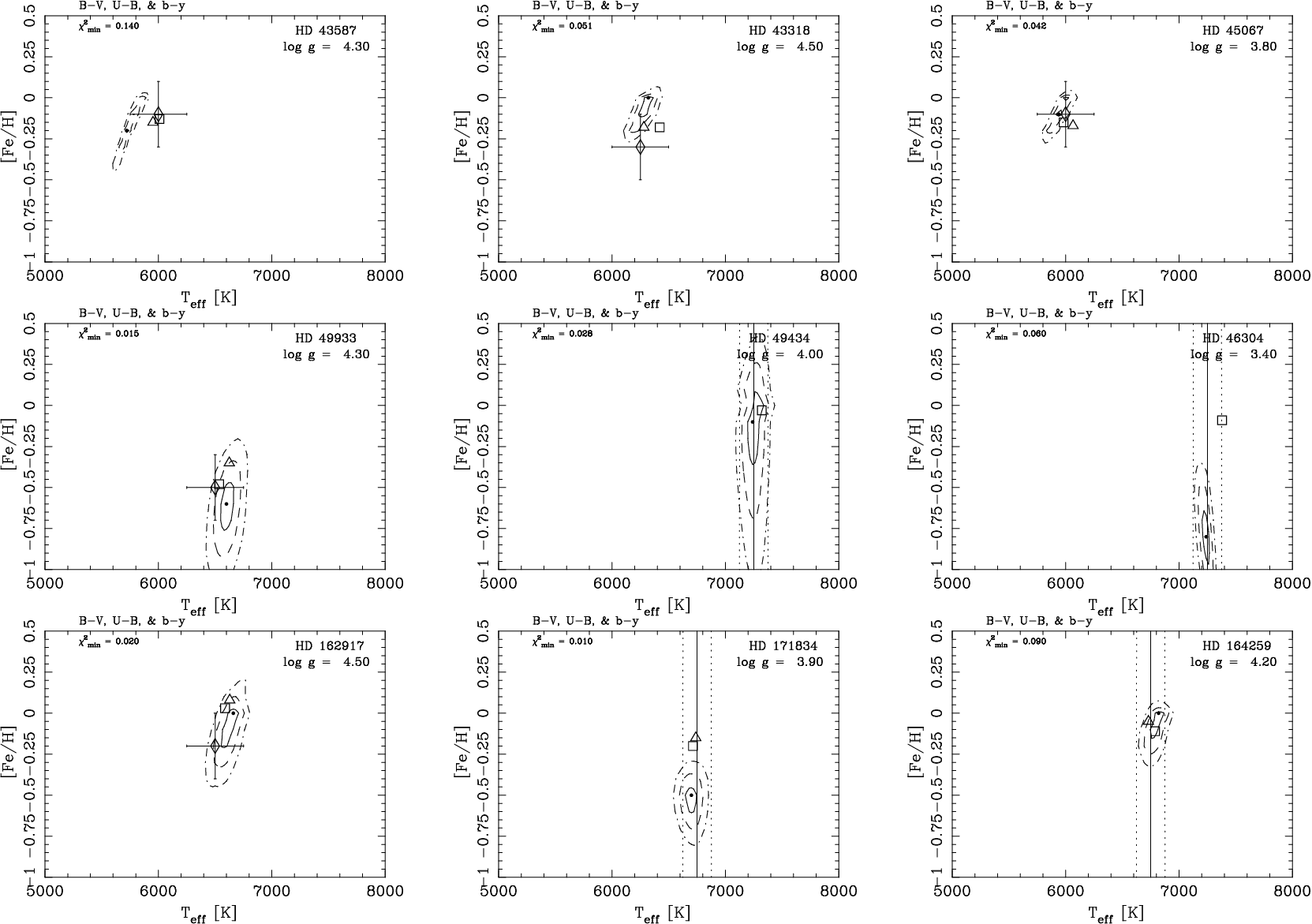,width=18.5truecm,height=19.6truecm,angle=-90.}}
\caption{As for Fig.~\ref{f:tefffeh_all}, but this time the solutions from the BaSeL models 
({\it 1-, 2- and 3-$\sigma$ confidence level contours}) are obtained in order to 
fit simultaneously the 3 following observed photometric values : (B$-$V),
(U$-$B),  (b$-$y). 
}
\label{f:final_tefffeh}
\end{figure*}

\begin{figure*}[htb]
\centerline{\psfig{file=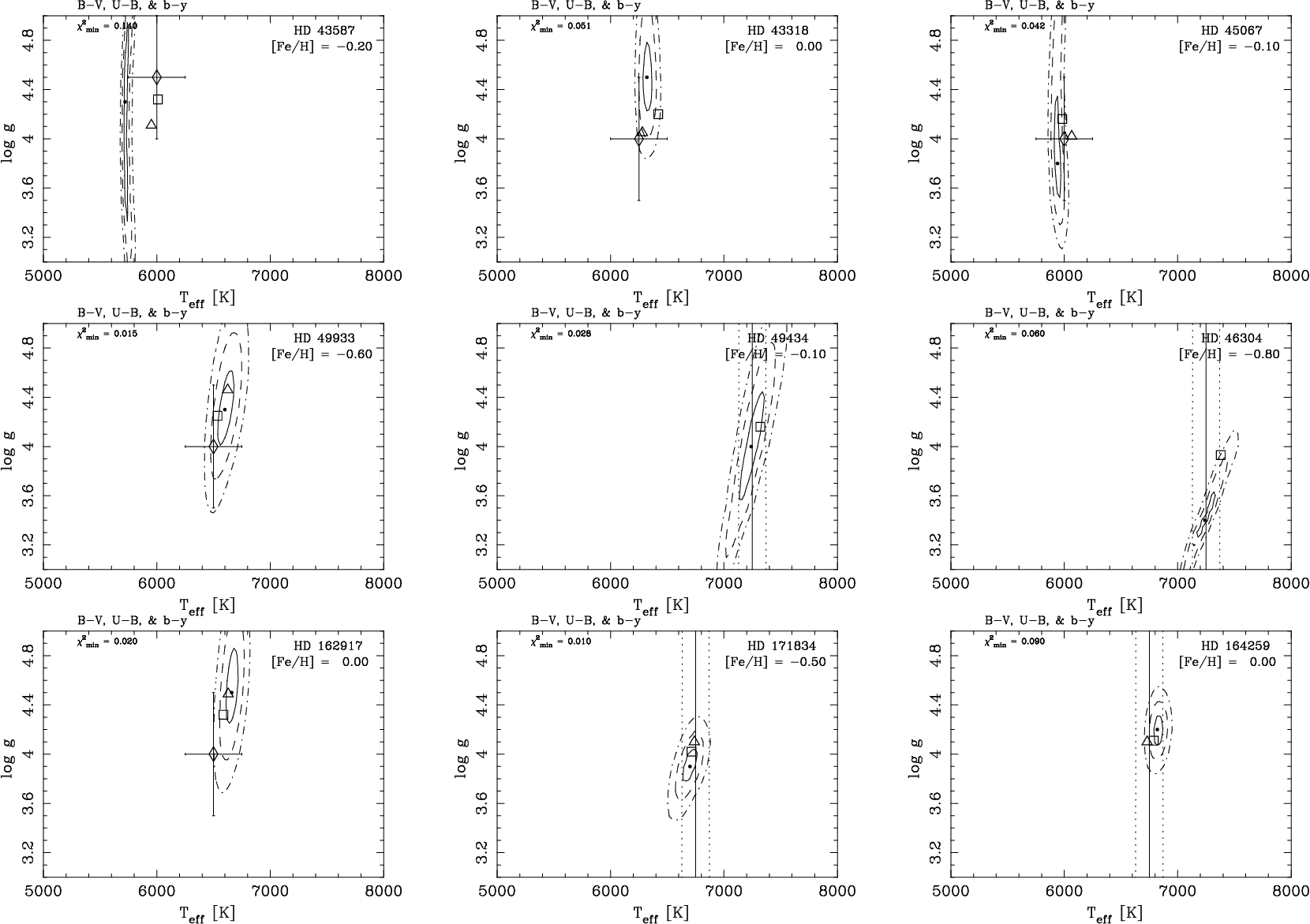,width=18.5truecm,height=19.6truecm,angle=-90.}}
\caption{
As for Fig.~\ref{f:tefflogg_all}, but this time the solutions from the BaSeL models 
({\it 1-, 2- and 3-$\sigma$ confidence level contours}) are obtained in order to 
fit simultaneously the 3 following observed photometric values : (B$-$V),
(U$-$B),  (b$-$y).  
}
\label{f:final_tefflogg}
\end{figure*}

Fig. \ref{f:final_tefffeh} and \ref{f:final_tefflogg} show the T$_{\rm eff}$-[Fe/H]-log g 
results by using the prescription suggested in the previous paragraph, namely, B$-$V, U$-$B, 
b$-$y. The results are also summarized in Table 2, where they are quoted as
BaSeL$^{c}$.  \\
A general comparison with the other methods shows that the BaSeL solutions are very satisfactory 
for the 3 fundamental parameters: 
effective temperatures are in excellent agreement, and [Fe/H] and gravities show good agreement. 
The improvement is clear when compared with the original results presented in Figs. 
\ref{f:tefffeh_all} and \ref{f:tefflogg_all} because there is no longer any systematic trend 
towards lower temperatures (see Tab. 2) and because the determinations of log
g and [Fe/H] are  also much better, without systematic deviations. \\
There is only one exception (HD 46304): its T$_{\rm eff}$ is in perfect agreement with 
its spectroscopic value, but the predicted log g and [Fe/H] are still low.  
While the BaSeL contours are consistent with the Templogg gravity only at the 2-$\sigma$ level, 
the metallicity predicted from the BaSeL models is poor in comparison to the result of the Templogg 
calibration ($\Delta$[Fe/H]$\simeq$0.7). 
What could explain this persistent difference ? 
It is worth noticing that this star has a large v sini (200 km s$^{-1}$, the
largest in our  working sample), and it is well known that high rotational
velocities modify the colours. Since the expected colour effect due to
rotation is typically a few hundredths of  a magnitude\footnote{These values
are highly dependent of spectral type, age, and chemical  composition, see for
instance Maeder (1971) and Zorec (1992).}  in B$-$V and increases with v sini,
this is probably part of the reason why the predictions of the  BaSeL models
disagree with the Templogg method for this star (even if rotation is not 
taken into account in the Templogg method).  \\ In conclusion, except in the
case discussed before, reliable and simultaneous estimates of the 3 
atmospheric parameters can be derived for F-type stars from only the three
synthetic BaSeL colours,  B$-$V, U$-$B and b$-$y.  This is a very useful
criterion for further applications. 

\section{Discussion}
Some of the intrinsic properties of the BaSeL library spectra can explain why synthetic 
Str\"om\-gren photometry does not increase the performance of UBV in determining 
fundamental stellar parameters (most particularly, abundan\-ces and surface gravities), 
as follows. 
\\
The fact that the library spectra have been calibrated using broad-band photometry 
implies that their use in determining fundamental stellar parameters from 
intermediate- and narrow-band photometry cannot be expected to provide results at 
the same level of confidence. 
Our results show that, in the particular domain studied, the broad-band 
calibration is adapted for the b$-$y index but not for the m$_1$ and c$_1$ indices. 
The reason is that, as b$-$y has been designed to measure the continuum, its synthetic 
value is less affected than m$_1$ and c$_1$ by high-resolution spectral features.  
Secondly, the library spectra have a typical resolution of 25\AA\, in the wavelength 
range where the corresponding synthetic photometry is being calculated. 
For the Johnson broad-band UBV (with passbands of half\-widths in the range 
500-1000\AA) a factor of $>$2 more flux data points are used in the numerical
integrations than are used in those for synthesizing the Str\"omgren data
(where passbands have halfwidths of order 200\AA). Since each flux point gives 
the integrated flux in a 25\AA\, passband and includes the effects of spectral 
lines and bands, the magnitude of these effects -- as well as their nonlinear
variations with abundance and/or surface gravity -- on the actually observed 
colours may be severely distorted in the computed colours (Buser 1978). 
As a consequence, the accuracy of synthetic Str\"omgren photometry may be 
significantly lower than it is for UBV, particularly for the m$_1$ and c$_1$ indices, 
which are weighted sums involving three passbands each (such that errors accumulate
faster than in UBV indices).   
\\
Finally, a further comment on the impressively good accuracy of the BaSeL 
determinations for the Johnson-Str\"om\-gren B$-$V, U$-$B, b$-$y combination 
(see Figs. \ref{f:final_tefffeh} \& \ref{f:final_tefflogg}) is in order. \\ 
As mentioned before, the Str\"omgren b$-$y index provides a reliable measure of the 
continuum and, therefore, a good temperature index. 
This is particularly interesting, given the known difficulties of matching empirical 
and theoretical B$-$V-T$_{\rm eff}$ scales to within better than about 0.03 mag 
(see, e.g., Sekiguchi \& Fukugita 2000)\footnote{
The b$-$y index is less vulnerable than B$-$V to the secondary effects of surface 
gravity and metallicity, even at the low spectral resolution given by the model spectra. 
}.
The fact that the U$-$B and B$-$V BaSeL colours give good measures of log g and [Fe/H] 
is not surprising. 
It is well known that if one does not ask UBV data to provide the temperature in the first 
place (e.g., by using an independent source, such as spectroscopy, spectral classification, 
or else suitable other photometry, such as Str\"omgren b$-$y), the sensitivities of 
both U$-$B and B$-$V can be used to full advantage for determining log g and [Fe/H]. 
Moreover, although these sensitivities change with temperature, they 
are near or even at their maxima in the F-dwarf star domain (e.g., Buser \& Kurucz 1992). 
This means that the derived values of [Fe/H] and log g presented in 
Figs. \ref{f:final_tefffeh} and \ref{f:final_tefflogg} are as reliable as they can possibly 
be, given the uncertainties in the colours. 

\section{Conclusion}
Several methods of determination of the fundamental stellar parameters T$_{\rm eff}$, 
log g and [Fe/H] are compared for nine single F stars. 
Particular attention has been paid to the simultaneous predictions of the BaSeL models 
in two photometric systems, Johnson and Str\"omgren.   
We show that using all photometric data is not the best strategy to obtain reliable 
simultaneous determinations of T$_{\rm eff}$, log g and [Fe/H] with the BaSeL models 
in the temperature range relevant for this paper, because of intrinsic limitations of 
the BaSeL library.
As a matter of fact, if one uses all five available photometric indices (B$-$V, U$-$B, b$-$y, 
m$_1$ and c$_1$), only the agreement with spectroscopic determination for the effective 
temperature is good, although the BaSeL-derived temperatures are slightly but systematically 
lower. The discrepancies in gravity and metallicity are however rather high, with the BaSeL 
predictions being too low. 
Alternatively, we show that the best results are obtained by using B$-$V, U$-$B, b$-$y 
in combination. 
This BaSeL combination is the best because on the one hand the b$-$y synthetic index 
gives reliable and accurate estimates of the effective temperature and, on the other hand, 
B$-$V and U$-$B give good estimates of [Fe/H] and the surface gravity. 
We wish to emphasize that, if Str\"omgren photometry is shown here to be less performing 
than UBV as a device for determining fundamental stellar parameters (and, most particularly, 
abundances and surface gravities), this is due to the fact that the BaSeL library spectra 
have been calibrated by colour-calibration in UBVRIJHKLM; it is not, of course, due to an 
intrinsic superiority of the UBV system over the Str\"omgren system. 
Thus, because the BaSeL library has only been calibrated in the UBVRIJHKLM colours, 
the best parameters derivations come from this system. 
Whilst parameters derivations in equivalent bandwidth systems such as Washington seem to be 
as good with the BaSeL libraries than with empirical methods (Lejeune 1997),  
parameter derivations from the BaSeL library in narrow band systems such as Str\"omgren 
photometry seem to be of poorer quality. \\
This work thus strongly suggests that the BaSeL library should be calibrated in various 
photometrical systems, in various bandwidth (maybe using higher resolution synthetic 
stellar spectra libraries) in order to represent in the best way the variety of the knowledge
of stellar spectral energy distributions that exists today in various photometric systems. \\ 
We found that the best combination to determine stellar parameters within the BaSeL 
library, even with its limitation, was to use a combination of two Johnson colours, 
U$-$B and B$-$V, and a Str\"omgren colour, b$-$y.\\
We also note that the agreement between Templogg and BaSeL for the hottest stars of the 
sample could be especially useful in view of the well known difficulty of spectroscopic 
determinations for fast rotating stars. \\
As far as the astrophysical applications are concerned, the BaSeL synthetic colours are 
of particular interest in evolutionary synthesis and colour-magnitude diagram studies 
of stellar populations, such as open clusters and young associations, where 
(i) F-type stars are highly common and, of course, 
(ii) a vast abundance of data are available in B$-$V, U$-$B and b$-$y colours.  
Concerning the determination of the atmospheric parameters of the COROT potential 
targets, a result of the present analysis is that all the methods give consistent solutions. 
In the context of the further $\sim$1000 potential targets of the COROT exploratory programme, 
it will be interesting to compare the results of the BaSeL models with those of 
automated spectral analysis methods (e.g. Katz {\al} 1998, and Bailer-Jones 2000).

\begin{acknowledgements}
We would like to thank J.-L. Vergely for providing extinction calculations
as well as J.C. Bouret, C. Catala and D. Katz for their participation in the
spectral caracterization of the stars. 
We are fully indebted to R. Kurucz for all the codes and theoretical data 
so generously displayed.  
E.L. and F.L. are supported by PPARC postdoctoral fellowship. 
Th. Lejeune gratefully acknowledges financial support from the  
Swiss NSF  (grant 20-53660.98 to R.B.), and Th. L\"uftinger
from the 'Fonds zur F\"orderung der wissenschaftlichen Forschung'. 
F.C. thanks FA\-PERJ for partial funding, through the grant 
E-26/171.368/1999. 
This research has made use of the QMW Starlink resource facilities 
and the SIMBAD database operated at CDS, Strasbourg, France. 
\end{acknowledgements}


\end{document}